\documentclass{PoS}
\usepackage[utf8]{inputenc}
\usepackage{amssymb}
\usepackage{amsmath}
\usepackage{amstext}
\usepackage{bbm} 
\usepackage{slashed} 
\usepackage{booktabs} 
\usepackage{csquotes} 
\usepackage{subcaption}
\newcommand{\dd}{\mathrm{d}}
\newcommand{\ii}{\mathrm{i}}
\newcommand{\e}{\mathrm{e}}
\newcommand{\tr}{\mathrm{tr}}

\title{$\mathcal{N}\!=\!1$ Supersymmetric $SU(3)$ Gauge Theory --\\ Pure Gauge sector with a twist}

\ShortTitle{$\mathcal{N}\!=\!1$ Supersymmetric $SU(3)$ Gauge Theory}

\author{\speaker{Marc Steinhauser}\\
        Friedrich Schiller University Jena, 07743 Jena, Germany\\
        E-mail: \email{marc.steinhauser@uni-jena.de}}

\author{Andre Sternbeck\\
        Friedrich Schiller University Jena, 07743 Jena, Germany\\
        E-mail: \email{andre.sternbeck@uni-jena.de}}
        
\author{Bj\"orn Wellegehausen\\
        Friedrich Schiller University Jena, 07743 Jena, Germany\\
        E-mail: \email{bjoern.wellegehausen@uni-jena.de}}
        
\author{Andreas Wipf\\
        Friedrich Schiller University Jena, 07743 Jena, Germany\\
        E-mail: \email{wipf@tpi.uni-jena.de}}

\abstract{Supersymmetric gauge theories are an
	essential part of most theories beyond the standard model. In the present work we investigate the pure gauge sector of Super-QCD focusing on the bound states, i.e. mesonic gluinoballs, gluino-glueballs and pure glueballs. To improve chiral properties and to minimize breaking of supersymmetry at finite lattice spacing, we introduce a deformed Super-Yang-Mills lattice action. It contains a twist term, similar to the twisted-mass formulation of lattice QCD. We furthermore explore if the multigrid method (DD$\alpha$AMG solver) applied to the gluinos (adjoint Majorana fermions) achieves similar improvements as in QCD.}

\FullConference{The 36th Annual International Symposium on Lattice Field Theory - LATTICE2018\\
		22-28 July, 2018\\
		Michigan State University, East Lansing, Michigan, USA.}

\begin{document}

\section{Introduction}

Supersymmetry is an ingredient of many extensions of the standard model of particle physics and has the potential
to solve open questions as for instance the hierarchy problem of the Higgs mass and the vacuum energy problem. 
In addition, the light superpartners of known
particles may serve as candidates for dark matter
in our universe. The research presented in these 
proceedings is focused on four-dimensional supersymmetric quantum chromodynamics (Super-QCD) \cite{Martin:1997ns,Wellegehausen:2018} and its essential building block, the $\mathcal{N}\!=\!1$ Super-Yang-Mills (SYM) theory with gauge group $SU(3)$. To study the non-perturbative features of this theory, we employ lattice simulations. 
Related investigations are performed by the DESY-M\"unster collaboration \cite{Ali:2018dnd,Gerber:2018,Lopez:2018}.

The on-shell $\mathcal{N}\!=\!1$ SYM theory contains a gauge field $A_\mu(x)$ and a Majorana field $\lambda(x)$ describing a gluon and its superpartner, the so-called gluino, in
interaction. Both fields are related by the supersymmetry transformation
\begin{equation}
	\delta_\epsilon A_\mu = \ii \bar{\epsilon}\gamma_\mu \lambda,\qquad
	\delta_\epsilon \lambda = \ii\Sigma_{\mu\nu} F^{\mu\nu} \epsilon
	\label{eq:SusyTrafo}
\end{equation}
with field strength tensor $F^{\mu\nu}$, infinitesimal anti-commutating constant Majorana spinor $\epsilon$ and
generators of the Lorentz-algebra  $\Sigma_{\mu\nu}\equiv\frac{\ii}{4}[\gamma_\mu,\gamma_\nu]$. The gluon field transforms in the adjoint representation and, as a consequence of supersymmetry, the fermionic gluino field transforms in the same representation. In the Minkowski spacetime the on-shell action reads
\begin{equation}
	S_\text{SYM} = \int \dd^4 x ~ \tr\left( -\frac{1}{4} F_{\mu\nu} F^{\mu\nu} + \frac{\ii}{2} \bar{\lambda} \slashed{D} \lambda - \frac{m}{2} \bar{\lambda}\lambda \right)\,.
	\label{eq:ContAction}
\end{equation}
For any finite gluino mass $m$ supersymmetry is broken softly. Such a term is however needed for lattice formulations of a supersymmetric gauge theory with Wilson fermions, e.g., as introduced by Curci and Veneziano \cite{Curci8612}. By including this gluino mass term, the counter-term of the symmetry breaking can be compensated via a fine-tuning of the bare gluino mass such that the renormalized gluino mass vanishes. Then, a continuum extrapolation will restore supersymmetry as well as chiral symmetry.

The particle spectrum of $\mathcal{N}\!=\!1$ SYM theory has been predicted by means of effective field theories based on symmetries as well as anomaly matching. Confinement requires that the states are color-neutral and supersymmetry leads to an arrangement of these states in supermultiplets, which are mass-degenerated as long as supersymmetry is unbroken. The first supermultiplet was predicted by Veneziano and Yankielowicz \cite{Veneziano8206} and consists of
\begin{table}[h]
	\centering
	\begin{tabular}{lll} \toprule
		1 bosonic scalar & $0^{++}$ gluinoball & $\text{a-}f_0\sim\bar{\lambda}\lambda$ \\
		1 bosonic pseudoscalar & $0^{-+}$ gluinoball & $\text{a-}\eta^\prime\sim\bar{\lambda}\gamma_5\lambda$ \\
		1 majorana-type spin $\frac{1}{2}$ & gluino-glueball & $~\,gg\,~\sim F_{\mu\nu}\Sigma^{\mu\nu}\lambda$ \\\bottomrule	
	\end{tabular}
\end{table}

The $SU(3)$ SYM action without a gluino mass term is invariant under a global chiral $U(1)_\text{A}$ symmetry \mbox{$\lambda\mapsto\exp(\ii\alpha\gamma_5)\lambda$}, but the anomaly reduces this to a $\mathbb{Z}_{6}$ symmetry
\begin{equation}
	\lambda\mapsto\e^{2\pi\ii n\gamma_5/6}\lambda\,,\qquad n\in\{1,\ldots,6\}\,.
	\label{eq:ChiralSym}
\end{equation}
A gluino condensate $\langle\bar{\lambda}\lambda\rangle\neq0$ spontaneously breaks this $\mathbb{Z}_6$ symmetry further to a $\mathbb{Z}_2$ symmetry. As a consequence the $SU(3)$ SYM theory has $N_\text{c}=3$ physically equivalent vacua \cite{Shifman:1987ia}.

\section{Lattice formulation with a twist}

To investigate the non-perturbative sector of $\mathcal{N}\!=\!1$ SYM theory, we perform lattice Monte Carlo simulations and estimate bound-state masses on a suitable number of gauge configurations. Various lattice formulations of the continuum on-shell action \eqref{eq:ContAction} exist, varying in the type of lattice fermions and the discretization of the continuum fields. Although all reasonable formulations lead to the same continuum limit, they differ at finite lattice spacing and how this limit is reached. To minimize the explicit breaking of chiral symmetry and supersymmetry we chose the lattice action $S_\text{lat}=S_\text{g}+S_\text{f}$ with the L{\"u}scher-Weisz gauge action
\begin{align}
	S_\text{g}[\mathcal{U}]&=\frac{\beta}{N_\text{c}} \bigg(\frac{5}{3}\sum_{\square}\tr(\mathbbm{1}-\text{Re}\,\mathcal{U}_\square) - \frac{1}{12}\sum_{\square\square}\tr(\mathbbm{1}-\text{Re}\,\mathcal{U}_{\square\square}) \bigg)\\
	\intertext{for the gauge boson and the fermionic action}
	S_{\text{f}}[\lambda,\bar{\lambda},\mathcal{U}]&=a^4\sum_{x,y\in\Lambda}\bar{\lambda}(x)D_\text{W}^{\text{tw}}(x,y)\lambda(y)
\end{align}
with the Wilson-Dirac operator
\begin{equation}
	D_\text{W}^{\text{tw}}(x,y)=(4+m+\ii\mu\gamma_5)\delta_{x,y} - \frac{1}{2}\!\sum_{\mu=1}^{4}\!\left\{\!\left( \mathbbm{1}-\gamma_\mu \right)\!\mathcal{V}_\mu(x)\,\delta_{x+\hat{\mu},y} + \left( \mathbbm{1}+\gamma_\mu \right)\!\mathcal{V}_\mu(x-\mu)^\dagger\delta_{x-\hat{\mu},y}\right\}\
\end{equation}
for the superpartner. Here $\mathcal{V}_\mu(x)$ denotes the gauge link in the adjoint representation, which one obtains from the traces $[\mathcal{V}_\mu(x)]^{ab}=2\,\tr[\mathcal{U}_\mu^\dagger(x)T^a \mathcal{U}_\mu(x)T^b]$ of the gauge link $\mathcal{U}_\mu(x)$ and the generator $T^a$ in the fundamental representation. 

Note the additional mass-like term $\ii\mu\gamma_5$, which is motivated as follows:
A non-zero mass $m$ favors a certain phase of the remnant 
discrete chiral symmetry \eqref{eq:ChiralSym}, which 
results in a non-vanishing gluino condensate $\langle\bar{\lambda}\lambda\rangle$. By introducing the additional mass-like term $\ii\mu\gamma_5$, a further gluino condensate $\langle\bar{\lambda}\gamma_5\lambda\rangle$ forms. These condensates are related by the $U(1)$ symmetry transformation in Eq.\,\eqref{eq:ChiralSym} and the \enquote{direction} of the condensate may be controlled via the parameters ($m$,~$\mu$). In contrast to QCD, the $\mathcal{N}\!=\!1$ SYM theory contains only one flavor and therefore the Wilson-Dirac operator does not contain the $\tau_3$ matrix as in twisted-mass QCD. Another difference lies in the treatment of observables: In the twisted-mass framework of lattice QCD, the twisted basis is back-rotated to the physical basis. We instead keep our basis fixed and use the parameter $\mu$ to deform our lattice theory. Although this deformation disappears when extrapolating to the chiral limit ($\mu\rightarrow0$ \& $m\rightarrow m_\text{crit}$), we expect that this additional parameter will be of advantage for the extrapolation. To this end, we approach the chiral point along a path where the bound states, which belong  to the Veneziano-Yankielowicz-supermultiplet in the continuum limit, have approximately equal mass  at finite lattice spacing. Specifically, we enforce that the chiral partners $\text{a-}\eta^\prime$ and $\text{a-}f_0$ have the same mass. In addition, we check if the degeneracy of $\text{a-}\eta^\prime$ and $\text{a-}f_0$ holds for the third state, the gluino-glue, as well.

In passing we note that a similar mass-like term has been successfully used in supersymmetric Wess-Zumino models on $2$-dimensional lattices \cite{Bergner:2007pu}, where the parameter $\mu$ (in the so-called non-standard Wilson term) has been tuned such that for free fermions the eigenvalues of the lattice Dirac operator reproduce those in the continuum up to $\mathcal{O}(a^4)$.

\section{Results}\label{ch:Results}

\begin{figure*}
	\begin{subfigure}[c]{0.33\textwidth}
		\includegraphics[width=\textwidth]{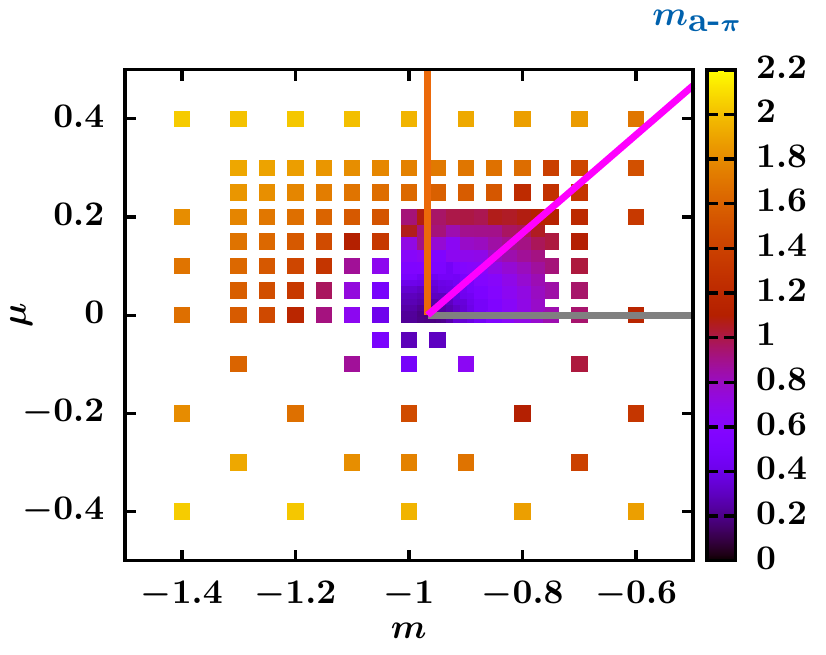}
	\end{subfigure}
	\begin{subfigure}[c]{0.33\textwidth}
		\includegraphics[width=\textwidth]{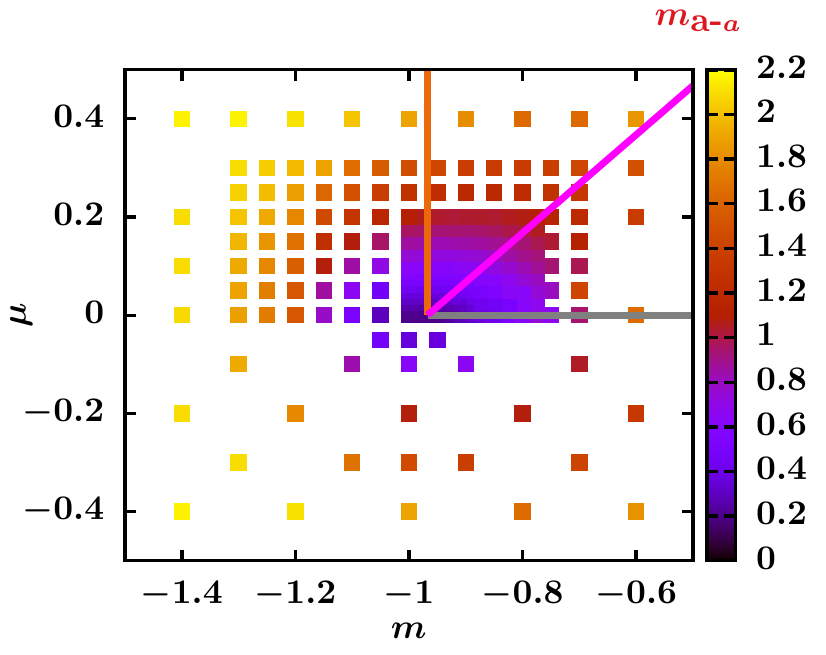}
	\end{subfigure}
	\begin{subfigure}[c]{0.33\textwidth}
		\includegraphics[width=\textwidth]{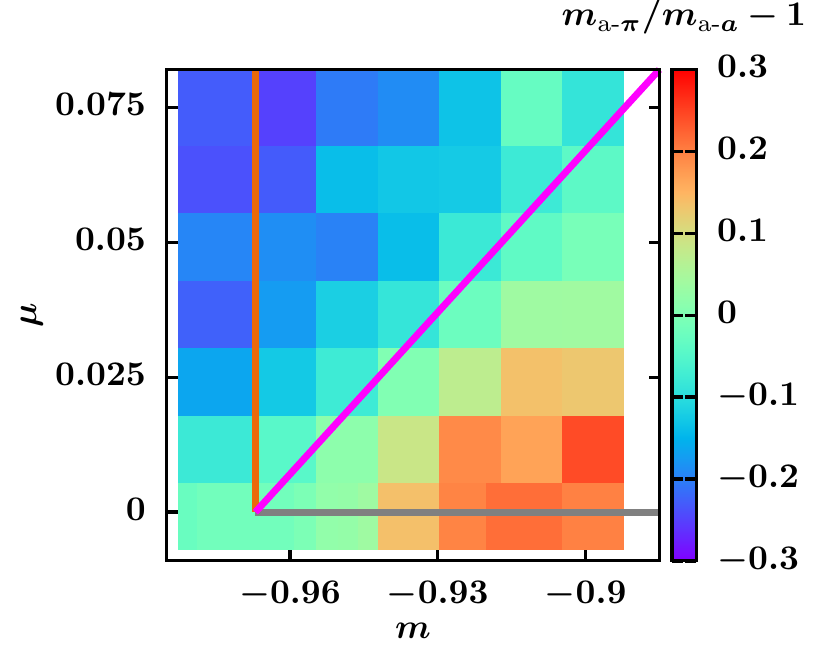}
	\end{subfigure}
	\caption{Masses of $\text{a-}\pi$ (left), $\text{a-}a$ (middle) and their subtracted ratio (right) in the $(m,\mu)$ plane on a $8^3\times 16$ lattice for $\beta=5.4$. The values are color-coded and the lines mark the three directions in $(m,\mu)$ mentioned in the text. Note the much smaller parameter range in the right plot.}
	\label{fig:PiAparameterScan}
\end{figure*}

First, we present our parameter scan in the $(m,\mu)$ plane for the gauge coupling $\beta=5.4$. We investigate the connected part of the $\text{a-}\eta^\prime$ correlator (called $\text{a-}\pi$) and the $\text{a-}f_0$ correlator (called $\text{a-}a$), which need much less statistics than the complete correlators. Figure~\ref{fig:PiAparameterScan} shows the masses of these states over a broad range of parameters on a $8^3\times16$ lattice. Plotting the subtracted ratio \hbox{$m_{\text{a-}\pi}/m_{\text{a-}a}-1$} in the vicinity of the chiral point ($m_\text{crit}=-0.967$, $\mu=0$) reveals three interesting directions which we parametrize by $\alpha\equiv\arctan\big(\mu/(m-m_\text{crit})\big)$:
\begin{center}
	\begin{tabular}{llll} \toprule
		$\alpha=\,0^\circ\,$ & $\mu=0$ & $m_{\text{a-}\pi}>m_{\text{a-}a}$ & gray line \\
		$\alpha=45^\circ$ & $\mu=m-m_\text{crit}$ & $m_{\text{a-}\pi}\approx m_{\text{a-}a}$ & magenta line \\
		$\alpha=90^\circ$ & $m=m_\text{crit}$ & $m_{\text{a-}\pi}<m_{\text{a-}a}$ & orange line \\\bottomrule	
	\end{tabular}
\end{center}
We see that if one approaches the critical point along the untwisted \mbox{$\alpha=0^\circ$} direction, the $\text{a-}a$-state is always lighter than the $\text{a-}\pi$-state. This changes for maximal twist \mbox{$\alpha=90^\circ$}, where the $\text{a-}a$ is always heavier than the $\text{a-}\pi$. For \mbox{$\alpha=45^\circ$} both masses are equal within errors, that is  $\text{a-}a$ and $\text{a-}\pi$ are approximately degenerated at finite lattice spacing.

\begin{figure*}
	\centering
	\includegraphics[width=0.95\textwidth]{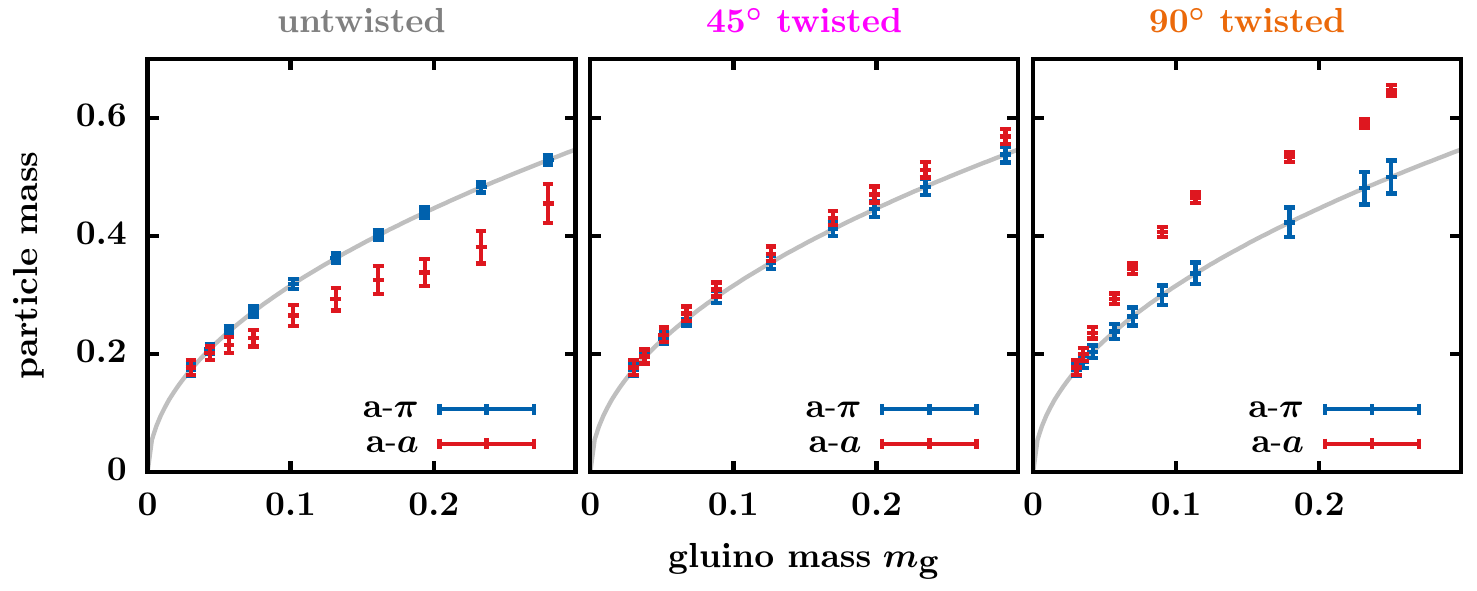}
	\caption{The masses of $\text{a-}\pi$ and $\text{a-}a$ for different twist angles $\alpha$ on a $8^3\times 16$ lattice for $\beta=5.4$. These plots are projections of the color-coded points along the three lines shown in Figure \protect\ref{fig:PiAparameterScan}.}
	\label{fig:PiAcuts}
    \medskip
	\includegraphics[width=0.95\textwidth]{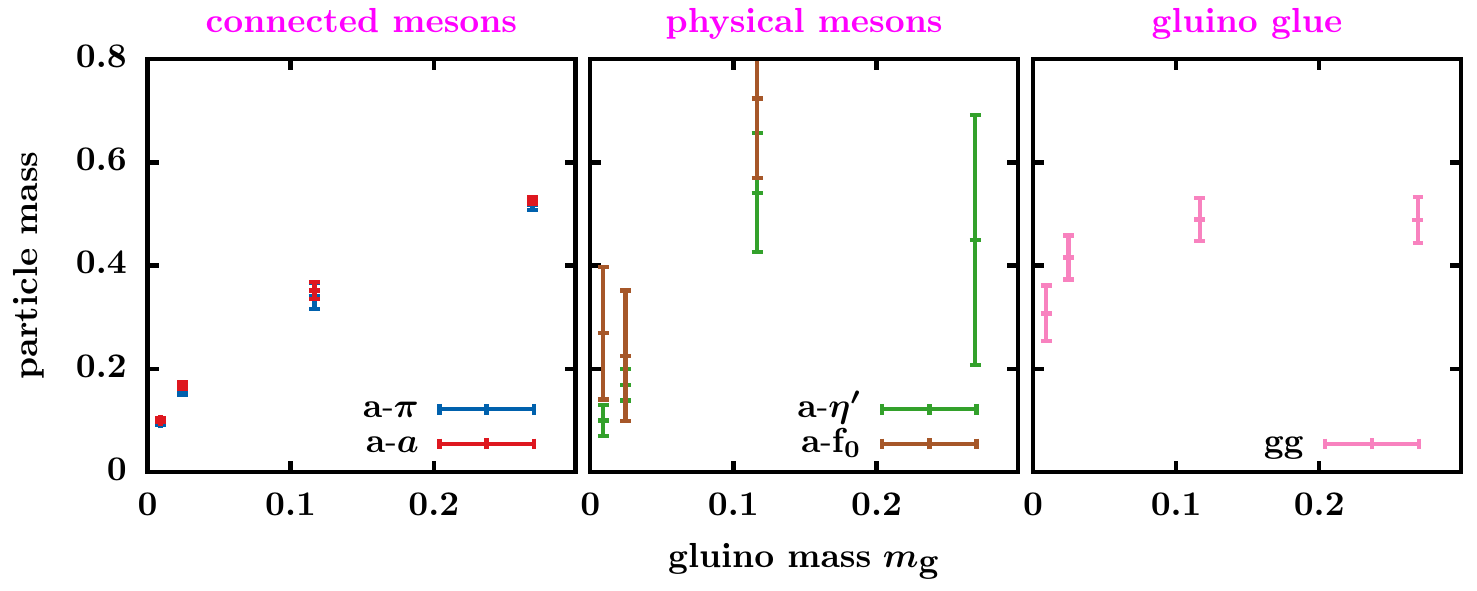}
	\caption{Masses of connected mesons $\text{a-}\pi$ \& $\text{a-}a$ (left), physical mesons $\text{a-}\eta^\prime$ \& $\text{a-}f_0$ (middle) and gluino-glueball $gg$ (right) for 4 different gluino masses on a $16^3\times 32$ lattice for $\beta=5.0$.}
	\label{fig:PiAEtaFgg16}
	\medskip\medskip\medskip
	\begin{minipage}{0.59\textwidth}
	Finally, we address the ubiquitous sign problem. The Pfaffian is part of the Boltzmann weight in the functional integral and hence should be positive. In general, \mbox{$\text{Pf}(D^\text{tw}_\text{W})\in\mathbb{C}$} for the twisted Wilson-Dirac operator. Nevertheless, in the continuum theory $m\rightarrow m_\text{crit}$, $\mu\rightarrow 0$, $a\rightarrow 0$ the Pfaffian becomes real. Our numerical investigation reveals that at finite lattice spacing the phase of \mbox{$\text{Pf}(D^\text{tw}_\text{W})=|\text{Pf}(D^\text{tw}_\text{W})|\cdot\e^{\ii\alpha}$} is negligible. Note that the computational costs for the calculation of the Pfaffian increases as $\mathcal{O}(N^3)$ with the matrix size $N$ and the memory requirement grows quadratically. Therefore the available computing resources limit us to a maximal lattice size of $7^3\times 14$. We show our measurements in Figure~\ref{fig:Pf} and extrapolate our data to $16^3\times32$, where the phase \mbox{$1-\cos(\alpha)<0.03$} indicates no severe sign problem.
	\end{minipage}\hfill
	\begin{minipage}{0.36\textwidth}
		\centering
		\includegraphics[width=0.9\textwidth]{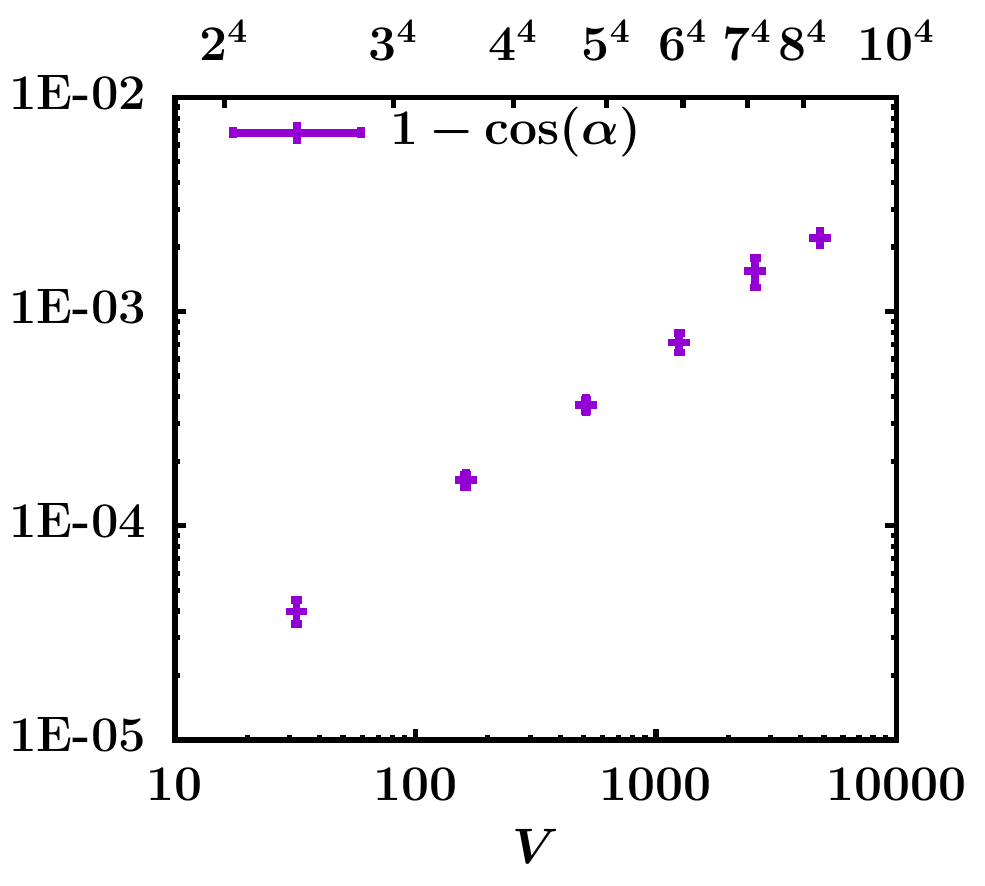}
		\caption{Measurement of the Pfaffian phase $\e^{\ii\alpha}$ plotted as \mbox{$1-\text{Re}(\e^{\ii\alpha})$} for the parameters \mbox{$m=-0.85$}, \mbox{$\mu=0.10$}, \mbox{$m_{\text{a-}\pi}\approx0.70$} on various lattice sizes.}
		\label{fig:Pf}
	\end{minipage}
\end{figure*}

For better visibility of the chiral extrapolation along the untwisted and two twisted directions, we plot in Figure~\ref{fig:PiAcuts} the masses $m_{\text{a-}\pi}$ and $m_{\text{a-}a}$ versus $m_\text{g}\sim m_{\text{a-}\pi}^2$. It is evident that a twist of \mbox{$\alpha=45^\circ$} is favored since the masses of the two superpartners are almost degenerate for this choice. Thus, in the following 
we focus on the $\alpha=45^\circ$ scenario. Further results at coupling $\beta=5.0$ (including the disconnected contributions to the mesonic states) on a $16^3\times 32$ lattice with approximately 500 configurations are shown in Figure~\ref{fig:PiAEtaFgg16}. The data of the mesonic states confirm that
the $\alpha=45^\circ$ scenario leads to the expected mass degeneracy.
Note that we have used Jacobi smearing to enhance the signal-to-noise ratio of all states. An adjustment of the smearing parameters and higher statistics are necessary for a better control of the uncertainties. For the gluino-glue instead the number of Jacobi smearing steps is less important but applying different levels of stout smearing improves the signal-to-noise ratio. The extracted masses are shown in the right plot of Figure~\ref{fig:PiAEtaFgg16}. We see these are comparable in size to the mesonic states, as expected for the supermultiplet in the continuum limit.


\begin{table}
	\centering
	\begin{tabular}{ccccc} \toprule
	lattice & $\beta$ & parameter $m$ & parameter $\mu$ & configurations \\\midrule
	$~8^3~\times16$ & $4.5$ & ~4~ values $\in[-0.968,-0.883]$ & 4 values $\in[0.007,0.092]$ & $~~~~4000$ \\
	$~8^3~\times16$ & $5.0$ & 14 values $\in[-1.061,-0.951]$ & 8 values $\in[0.000,0.085]$ & $~~~~4000$ \\
	$~8^3~\times16$ & $5.4$ & 28 values $\in[-1.400,-0.600]$ & 19 values $\in[-0.400,0.400]$ & $~~~~~~200$ \\
	$~8^3~\times16$ & $5.4$ & ~4~ values $\in[-1.217,-1.168]$ & 4 values $\in[0.007,0.057]$ & $~~~~4000$ \\
	$16^3\times32$ & $5.0$ & ~4~ values $\in[-1.064,-1.011]$ & 4 values $\in[0.007,0.060]$ & $\sim1000$
	\\\bottomrule
	\end{tabular}
	\caption{Overview of simulation parameters.}
\end{table}

\section{DD$\alpha$AMG}

\begin{figure}
	\centering
	\includegraphics[width=\textwidth]{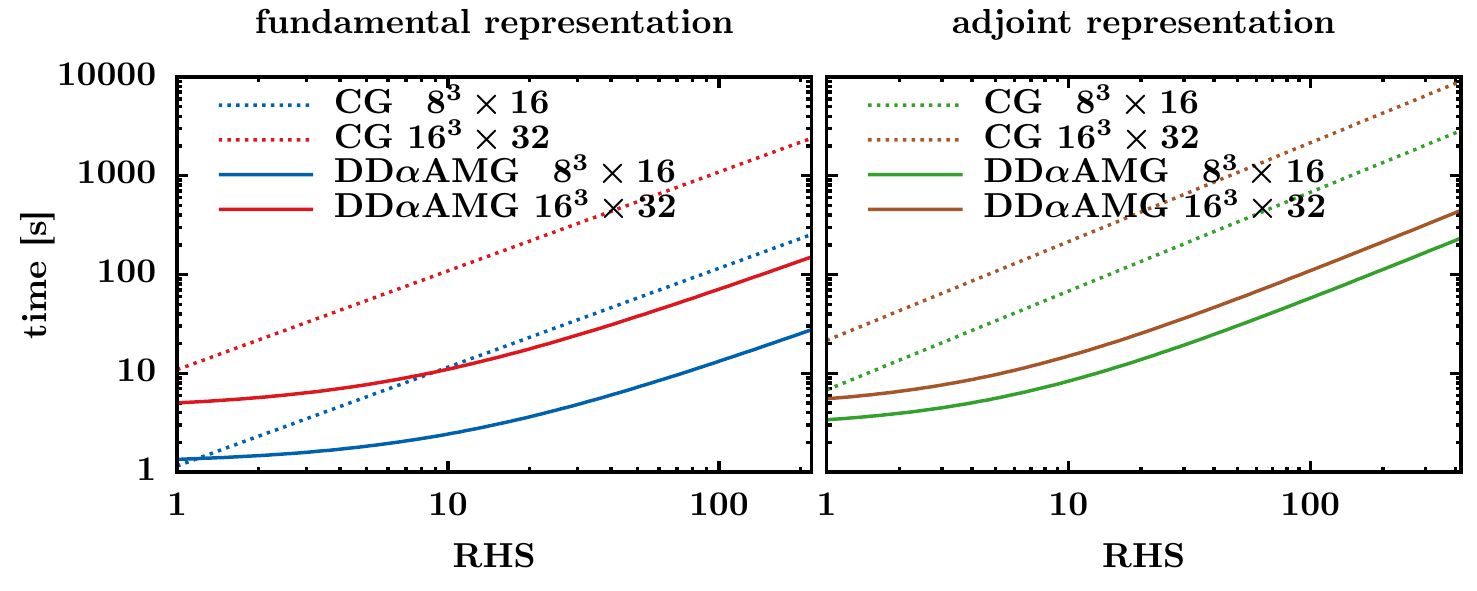}
	\caption{Time to solution for different numbers of right hand sides (RHS). The left (right) figure compares the timing for the fundamental (adjoint) representation of a standard CG solver with the DD$\alpha$AMG solver on lattices of size $8^3\times16$ and $16^3\times32$. Note the offset due to the DD$\alpha$AMG setup which becomes negligible for a large number of RHS.}
	\label{fig:DDalphaAMGbenchmarks}
\end{figure}

Many stochastic estimators are required for the measurement of the $\text{a-}\eta^\prime$ and $\text{a-}f_0$ correlators. This computation requires many inversions $x=D^{-1}y$ of the Wilson-Dirac operator, which can be accelerated with the help of a multigrid algorithm. We used the adaptive aggregation-based domain decomposition multigrid (DD$\alpha$AMG) library \cite{Alexandrou:2016izb,DDalphaAMG}. This solver is based on two ingredients: On the one hand, the Schwarz alternating procedure (SAP) utilizes domain decomposition and deals with the UV-modes. On the other hand, the coarse grid correction is an interpolation operator which approximates the small eigenvalues to tackle the IR-modes. We generalized the hard-coded $SU(3)$ gauge group in the library to $SU(N_\text{c})$ with arbitrary $N_\text{c}$ and representation.
For benchmarks we choose the following setting: gauge group $SU(3)$ in the fundamental and adjoint representation, lattice sizes $8^3\times16$ and $16^3\times32$, two multigrid levels, block size $2^4$, mixed precision and solver combination FGMRES + red-black Schwarz. Figure~\ref{fig:DDalphaAMGbenchmarks} shows the timings for up to 100 stochastic estimators and 5 point sources and a comparison to an ordinary conjugate gradient (CG) algorithm. For these scenarios we achieve a speed-up factor of $9$ to $20$.

\section{Summary}

Our numerical study of the $\mathcal{N}\!=\!1$ Super-Yang-Mills theory introduces two novel concepts: Firstly, we deform our lattice formulation with Wilson fermions by introducing an additional 
mass-like term, similar to a one-flavor formulation of twisted-mass QCD. By tuning the two mass parameters of our lattice action, we achieve a considerably improved mass degeneracy of the chiral partners in the Veneziano-Yankielowicz supermultiplet at finite lattice spacing. Furthermore, we find that the superpartner, the so-called gluino-glue, has a similar mass at finite lattice spacing. Our results suggest that with the deformed action a faster convergence towards the continuum-limit may be achieved.

In addition, we adapted the DD$\alpha$AMG library for arbitrary representations of the gauge group $SU(N_\text{c})$ and used it to accelerate the inversion of the Wilson-Dirac operator for the bound-state mass measurements. Benchmarks show an impressive speed-up of about 20 for the adjoint $SU(3)$.
\newline\newline
\textbf{Acknowledgments:}
We thank Stefano Piemonte and Sara Collins for helpful comments and the Leibniz Supercomputing Centre for funding the project pr48ji. M.S.\ received the student support of the conference and A.S.\ was supported by the BMBF under grant No.\ 05P15SJFAA.

\bibliographystyle{JHEP}
\bibliography{references}

\end{document}